\documentstyle[12pt]{article}
\begin{document}

\newcommand{\tr}{\bigtriangleup}
\newcommand{\rv}{\stackrel{\leftarrow}
{\frac{\delta }{\delta \eta}}}
\newcommand{\rvs}{ * \stackrel{\leftarrow}
{\frac{\delta }{\delta \eta}}}
\newcommand{\vv}[2]{\frac{\,\delta #1}{\delta \eta #2 }}
\newcommand{\vvv}[3]{\frac{\delta^2 #1}{\delta \eta #2
                     \,\delta \eta #3 }}
\newcommand{\bin}[2]{\left(\!\!\begin{array}{c}#1\\#2
                     \end{array}\!\!\right)}
\newcommand{\free}{ m^2 - \partial\,^2}
\newcommand{\vgN}{\{{\cal G}^N\}}
\newcommand{\lb}{\lambda}
\newcommand{\be}{\begin{equation}}
\newcommand{\ee}{\end{equation}}
\newcommand{\ba}{\begin{eqnarray}}
\newcommand{\ea}{\end{eqnarray}}

\title{Schwinger equation as singularly perturbed equation}

\author{V.E. Rochev  and P.A. Saponov \\
\small\it Institute for High Energy Physics,
142284 Protvino, Moscow region, Russia}
\date{}
\maketitle

\begin{abstract}
A new approximation scheme for non-perturbative
calculations in a quantum field theory is proposed. The scheme is
based on investigation of solutions of the Schwinger equation
with its singular character taken into account. As a necessary
supplementary boundary condition the Green functions' connected
structures correspondence principle is used. Besides the usual
perturbation theory expansion which is always available as a
particular solution of our scheme
some non-perturbative solutions of an equation for the propagator
are found in the model of a self-interacting scalar field.
\end{abstract}

\noindent
{\large \bf Introduction}
\vskip 2 true mm

As is well known there exists a wide class of physical phenomena
(called non perturbative effects) which cannot be described
by a finite number of terms of the perturbation theory series.
It is usually believed that the full or partial summation of
the perturbative series solves this problem. Such an approach
supposes implicitly that the sum of the perturbative series
contains exhaustive information about a given quantum field
model.

Meanwhile even simple mechanical models provide an example
for the perturbative series to be unable to describe a physical
situation properly.

Thus, it is well known from fluid mechanics that one cannot
use an ideal liquid as the leading approximation to a viscous one
(even if the  viscosity is very small) when considering a process
of flowing near the boundary of an  immersed body. It is a
consequence of the fact that the viscous liquid is a so called
singularly perturbed system \cite{1,2} compared to the ideal
liquid. A solution of equations of motion for the ideal liquid
as well as the perturbation theory based on this solution cannot
obey the boundary conditions for the viscous liquid in principle.

A characteristic feature of a singularly perturbed system is
existence of an essential singularity with respect to an expanding
parameter in  solutions. If such a system is described by a
differential equation then the highest derivative coefficient depends
on an expanding parameter as follows:  when this parameter is equal
to zero the order of the equation decreases and the number of
necessary boundary conditions reduces. This means that the
perturbative series is a solution of the initial problem only for
some special boundary conditions. If a physical situation goes beyond
their framework the result of the perturbation theory summation fails
to describe such a situation in principle.

Let us consider an elementary example. Take the equation for a
system in the form:
\be
\lb\,{\dot x} = t - x \label{1}
\ee
with the boundary condition $x(0) = X$.
At $\lambda =0$ the order of the equation lowers so (\ref{1}) is a
singularly perturbed equation. In this case the perturbative series
on $\lambda$ breaks at the second term and the result of its
summation is
\be
x_{pert}\,(t) \equiv \sum \lb^n x_n = t - \lb.
\label{2}
\ee
Though $x_{pert}$ is an exact solution of (\ref{1}) with the special
(perturbative) boundary condition $x_{pert}(0)=-\lambda$, it cannot
serve as a solution of the problem (\ref{1}) when $X\not=-\lambda$.
The exact solution of (\ref{1}) has the form
\be
x(t) = t - \lb + (X + \lb)\,\exp (- \frac{t}{\lb}) \label{3}
\ee
and shows that in solving a problem with non-perturbative boundary
conditions by iterations one should introduce a so called boundary
series besides the usual perturbative one. In our simple example
(\ref{1}) this boundary series consists of the two terms containing
an essential singularity on $\lambda$.

The boundary series contribution cannot be taken into account by
perturbation theory and in general this contribution dominates near
the point $t=0$ that is outside the region of the Tikhonov's theorem
applicability \cite{2}. This domination can be easily seen from our
example if we calculate the derivatives of $x(t)$ at  $t=0$. We get
${\dot x}(0)=-{1 \over \lb}X,\,{\ddot x}(0)={1 \over \lambda^2}X +
{1 \over \lambda}$ and
so on. These derivatives are of obviously non-perturbative
character (except for the case of the perturbative boundary
condition $X=-\lambda$).

Now let us turn to the case of quantum field theory. An essential
singularity with respect to a coupling constant is the wide known
attribute of any field theory model with interaction. Moreover,
the Schwinger equation for the generating functional contains a
coupling constant as a coefficient at the highest functional
derivative so from the point of view of the differential equations
theory the Schwinger equation is a typical example of a singularly
perturbed equation.

It is also worth mentioning that it is Green functions (vacuum
expectation values) which are of physical interest. As these
functions are the derivatives of the generating functional at the
source switched off (hence, in the obviously non-Tikhonov region)
so the boundary series contribution to Green functions will dominate
except for the perturbative boundary conditions. In this work we
make an attempt to treat the Schwinger equation as singularly
perturbed one and to construct a new approximation scheme of solving
of this equation.

The proposed scheme  approximates the non-perturbative Green
functions already at first step without summing of expansions and at
the same time contains the standard perturbation theory as a
particular case. The scope of admissible boundary conditions becomes
wider and they do not limited by the framework of the perturbation
theory any more. It opens new possibilities for describing of
essentially non-perturbative effects (such as mass spectrum
reconstruction) already at the leading orders of approximation.
We hope that the proposed approximation scheme  (or those similar to
it) can prove to be useful when studying the problems connected with
the spontaneous symmetry breaking and also in problems where one
should take into account from the very beginning the nontrivial
properties of the physical vacuum as media.

Our work is organized as follows. In part~1 necessary definitions are
introduced and general principles of the approximation scheme
construction are stated. The primary version of the scheme \cite{3}
based on a general solution of the Schwinger equation is considerably
simplified and in the form proposed here this scheme can be
generalized to any quantum field model (in present work  we deal with
the simplest case of a self-interacting scalar field only). In part~2
the equation for the propagator at the leading approximation is
solved. Brief discussion of the results contains in Conclusion.
\vskip 5 true mm

\noindent
{\large \bf 1. Schwinger Equation and Approximation Scheme}
\vskip 2 true mm

We will consider the theory of complex scalar field $\phi (x)$ in
$d$-dimensional Euclidean space $x\in E_d$ with nonlocal
self-interaction:
\be
S_{int} = \int dx\,dy\,(\phi^*\phi)\,(x)\,\frac{\lb\,(xy)}{2}\,
(\phi^*\phi)\,(y). \label{4}
\ee
The limit $\lambda (xy)\rightarrow \lambda\delta (x-y)$ corresponds
to the local theory with the quadric interaction.

To define the Green functions we introduce a bilocal source
$\eta (xy)$\footnote{The use of the bilocal source is made only for
the sake of simplicity and compactness of calculations and is not a
principal point in proposed formulation of the approximation
scheme.}. Then the $n$-particle (2n-point) functions (vacuum
expectation values) are the derivatives of the generating functional
\ba
G &=& \sum_{n=0}^{\infty} \frac{(- 1)\,^n}{n!} G_n \eta^n =
\nonumber\\
  &=& \sum_{n=0}^{\infty} \frac{(- 1)\,^n}{n!} \int
      \prod_{k=1}^{n}dx_k\,dy_k\,G_n\!\bin{x_1\dots x_n}
      {y_1\dots y_n}\,
      \eta(y_1 x_1)\dots \eta(y_n x_n).
\label{5}
\ea
The first derivative is the propagator of the particle:
\be
\tr(x y) = \left. G_1\!\bin{x}{y} = - \vv{G}{(y x)}\,
           \right|_{\eta=0},\label{6}
\ee
the second one is the two-particle function:
\be
G_2\!\bin{x_1\,x_2}{y_1\,y_2} =\left. \vvv{G}{(y_2 x_2)}{(y_1 x_1)}\,
\right|_{\eta=0}.\label{7}
\ee

The Schwinger equation for the generating functional is a
consequence of the field equations and the quantization conditions.
For the model involved it has the form:
\ba
& &\int dx'\,\lb\,(x x')\,\vvv{G}{(y x')}{(x' x)} = \nonumber\\
& &(\free)\,\vv{G}{(y x)} + \int dx'\,\eta(x x')\,\vv{G}{(y x')}
+ \delta(x - y)\,G.\label{8}
\ea
Here $m^2$  stands for the mass of the free field $\phi(x)$ and
$\partial^2\equiv \sum^{d}_{n=1}\frac{\partial^2}{\partial x^2_n}$
is the Laplace  operator.

The following notations will be subsequently used:
\ba
( AB )\,(x y) & \equiv & A(x y)\, B(x y), \quad
\vv{}{}(x y)  \equiv \vv{}{(y x)}, \nonumber \\
( A*B )\,(x y)&\equiv& \int dx' A(x x')\,B(x' y).
 \label{9}
\ea

In these  notations the Schwinger equation (\ref{8}) reads as
follows:
\be
\left(\lb\,\vv{G}{}\right)\rvs = (\free + \eta\,) * \vv{G}{}
+ G.\label{10}
\ee

Successive differentiation of (\ref{10}) gives (on switching off
the source) the infinite system of the Dyson equations for the Green
functions $G_n$.

The Schwinger equation (\ref{10}) belongs to the class of singularly
perturbed equations. Indeed, when $\lambda =0$ the order of the
equation reduces: the generating functional $G^{(0)}$ at the leading
order of the perturbation theory on $\lambda$ (that is the free
field generating functional) is a solution of the first order
equation:
\be
(\free + \eta\,) * \vv{G^{(0)}}{} + G^{(0)} = 0.\label{11}
\ee
The solution of (\ref{11}) is uniquely fixed by the only boundary
condition: the normalization of the generating functional
\be
G\,[\eta = 0] = 1.\label{12}
\ee
This solution is:
\be
G^{(0)} = \det |\,(\free + \eta\,)^{-1} * (\free)\,|.\label{13}
\ee

The perturbative solution for the Schwinger equation
\be
G_{pert} = \sum_{n=0}^{\infty} G^{(n)},\label{14}
\ee
where $G^{(n)} = O(\lambda^n)$ is constructed by the iterations. The
equation for $G^{(n)}$ is again of the first order with respect to
the functional derivatives
\be
(\free + \eta\,)\,*\,\vv{G\,^{(n)}}{} + G\,^{(n)} =
\left(\lb\,\vv{G\,^{(n-1)}}{}\right)\rvs \label{15}
\ee
and therefore to fix $G^{(n)}$ at any order of the perturbation theory
on $\lambda$ one needs the only condition $G^{(n)}[0]=0$ which is a
simple consequence of the normalization condition (\ref{12}).

Meanwhile at $\lambda\not=0$ (even if small) the Schwinger equation
(\ref{10}) is that of the second order with respect to the functional
derivatives and for its solution to be fixed uniquely one needs a
supplementary boundary condition. The problem of boundary conditions
is understood to be the key one for constructing of non-perturbative
solutions of the Schwinger-Dyson equations  \cite{4,5,6}.

Being the  solution of the Schwinger equation (\ref{10}), the
perturbative series (\ref{14}) has the doubtless advantage that there
is no need for fixing a supplementary boundary condition. But this
advantage can turn to be a serious lack: physical phenomena described
in principle by a given field model can escape out of the
consideration. The perturbation theory even being summed up can fail
to obey the physical boundary conditions just as the ideal liquid is
unable to provide the physical boundary conditions for the viscous
liquid in the boundary layer. At the same time enlarging of the class
of admissible boundary conditions can provide the description of
non-perturbative phenomena after making a finite number of iterations
and without summing of the corresponding expansions.

Let us turn to the construction of the approximation scheme. Since
the perturbation theory is the only universal tool for calculating
of the Green functions in quantum field theory we will use it as a
base that is will construct our scheme in such a way that the
perturbation theory expansion would be contained in it as a
particular case.

Introduce a perturbative approximant:
\be
G_{pert}^N \equiv \sum_{n=0}^N G^{(n)}\label{16}
\ee
and define a non-perturbative approximant
$G^N$ to be found  as a solution of the equation:
\ba
\left(\lb\,\vv{ G^N}{}\right)\rvs \!&-&\! (\free + \eta\,) *
\vv{G^N}{} - G^N =\nonumber \\
\left(\lb\,\vv{G_{pert}^N}{}\right)\rvs\! &-&\! (\free + \eta\,) *
\vv{G_{pert}^N}{} - G_{pert}^N.
\label{17}
\ea
When $N\rightarrow\infty$ the perturbative approximant $G_{pert}^N$
tends to the perturbative solution of the Schwinger equation
$G\,_{pert}$ and therefore $G^N$ tends to some exact solution of the
same equation with, in general, arbitrary boundary
conditions\footnote{We do not discuss the problem of convergency of
expansions.}.
That is why it would be more correct to speak about a set of
approximants $\{G^N\}$ every element of which obeying definite
boundary conditions. Under $N\rightarrow\infty$ this set of
approximants turns into the full set of the all solutions of the
Schwinger equation. $G_{pert}^N\in \{G^N\}$ as the trivial
solution of equation (\ref{17}).

To construct a solution of equation (\ref{17}) one needs to formulate
some criteria restricting the framework of the admissible
supplementary boundary conditions so as in general case they are
arbitrary. After such a restriction we obtain a certain subset
$\{{\cal G}^N\}$ of the all possible solutions of equation (\ref{17}):
$\{{\cal G}^N\}\subset \{G^N\}$. We will choose the boundary
conditions in such a way (see below) that the set $\vgN$ would
include $G^N_{pert}$. Our scheme will be nontrivial if it will
contain other solutions too. In such a case one comes again to the
question of the choice of the proper solution for the given model
among the elements of $\vgN$. This choice can be made on taking into
account some additional physical requirements. One of them is the
principle of minimal energy value of the ground state.

So, summing all mentioned above, our goal is to reduce the full set
of the all possible solutions $\{G^N\}$ to some subset $\vgN$ by
imposing additional boundary conditions on the solutions of equation
(\ref{17}).

In formulating such conditions we will base ourselves on those
properties of the perturbative approximant which can be generalized
to the non-perturbative case. In so doing we extremely restrict the
admissible boundary conditions but the perturbative solution is still
contained in $\vgN$.

One of the properties mentioned above is the connected structure of
the perturbative approximant. General connected structure of Green
functions can be found beyond the framework of the perturbation
theory with the help of the theorem on the connectivity of the
logarithm\footnote{It is worth mentioning that to find the connected
structure one should introduce the simple sources. The derivatives
of $Z=\ln G$ with respect to the bilocal source are not in general
the connected functions.}. For example, the connected structure of
the two-particle function is:
\be
G_2\!\bin{x_1\,x_2}{y_1\,y_2} = \tr\,(x_1\,y_1)\,\tr\,(x_2\,y_2)
+ \tr\,(x_1\,y_2)\,\tr\,(x_2\,y_1) + G_2^{con}\!\bin{x_1\,x_2}
{y_1\,y_2},
\label{18}
\ee
$G_2^{con}$ being the connected part of the two-particle function and
$\tr$ is the full propagator. For the three-particle function this
structure is of the form
\be
G_3 = \mbox{Sym}\left\{ 6\tr \tr \tr + 9\tr G_2^{con}\right\}
+ G_3^{con} \label{19}
\ee
where the notation Sym stands for the Bose-symmetrization. Formulae
(\ref{18} -\ref{19}) (and those similar to them for higher functions)
are valid not only at any order of the perturbation theory but also
beyond its frames.

At $N$-th order of the perturbation theory all the connected parts
of $n$-particle functions  are equal to zero when $n>N+1$. For
example, at the leading order only the propagator is not equal to
zero, at the order $O(\lb)$ $(N=1)$ the propagator and the connected
part of the two-particle function are non-vanishing, and so on.

To say it another, dynamics at $N$-th approximation is fully defined
by $N+1$ lowest Green functions. In this connection at $N$-th order
one can confine oneself to considering the subsystem of $N+1$
equations from the whole infinite Dyson system. So as at
$N\rightarrow\infty$ the $N+1$ equations turn into the full system
then the truncated approximant (a solution of $N+1$  equations) still
tends to the solution $G_{pert}$.

We lay these properties into the base of our scheme construction.
Physically this means that we consider the non-perturbative solutions
of the Schwinger equation of such a type that the highest Green
functions weakly affect dynamics defined by the lowest functions that
is, for example, one can neglect the three-particle forces when
considering the leading approximation to the two-particle processes.

Thus, to construct the $N$-th step of the approximation scheme
one should proceed in the following way:

1. At the $N$-th step we consider $N+1$ equations of the Dyson type
following from (\ref{17}) that is equation (\ref{17}) itself and its
$N$ derivatives at $\eta=0$.

2. Due to $N+1$ equations contain $(N+2)$ $n$-particle functions
$G^N_1,...,G^N_{N+2}$ this system needs a supplementary boundary
condition\footnote{Normalization condition (\ref{12}) is supposed
to be fulfilled at any step of consideration.}. To introduce such a
condition we will demand the correspondence between the connected
structures of the sought approximant and the perturbative one.
This perturbative approximant consists of the set of $(N+2)$
$n$-particle functions calculated with the perturbation theory
which will be denoted as
$$
g^N_n \equiv (-1)^n\,\left. \vv{^n G_{pert}^N}{\,^n}\right|_{\eta=0}.
$$

For the perturbative approximant $[g^N_{N+2}]^{con}=0$ therefore
we put for our system
\be
\left[{\cal G}^N_{N + 2}\right]^{con} = 0.\label{20}
\ee

The connected structures correspondence condition (\ref{20}) allows
one to express $(N+2)$-particle function in terms of $N+1$ lowest
ones
\be
{\cal G}^N_{N +2} = {\cal G}^N_{N + 2}\left[{\cal G}^N_{1},
\dots {\cal G}^N_{N + 1}\right]
\label{21}
\ee
and by this to close the system of equations.

At every step of the approximation scheme we get, generally speaking,
some collection of the $n$-particle functions sets
$\{{\cal G}^N_1$ ,...,$ {\cal G}^N_{N+2}\}$. The set of perturbative
$n$-particle functions $\{g^N_1,...,g^N_{N+2}\}$ is also contained
in the collection. {\it A priori} the region of the scheme
applicability is that of the small $\lb$.
 When $N\rightarrow\infty$ each of these sets tends
to the full (infinite) number of Green functions corresponding to some
exact solution of the Schwinger equation (\ref{10}). Each of these
solutions has the true connected structure.

Let us elaborate the two primary steps in more detail. For $N=0$ the
perturbative approximant $G_{pert}^0=G^{(0)}$ is the generating
functional of the free Green functions (\ref{13}). The system
consists of one equation --- this is equation (\ref{17}) at $\eta=0$:
\ba
\int dx'\,\lb\,(x x')\,{{\cal G}\,}^0_{2}\!\bin{x\,x'}{y\,x'} +
(\free)\! & &\mbox{\hspace*{-6 true mm}} \tr^0(x\,y) - \delta(x - y)
\nonumber\\
=& &\mbox{\hspace*{-6 true mm}} \int dx'\, \lb\,(x\,x')\,
g\,^0_{2}\!\bin{x\,x'}{y\,x'}.
\label{22}
\ea
Here
\be
\tr^N = - \left. \vv{{\cal G}^N}{}\right|_{\eta = 0},
\ee
and
\be
g^0_{2}\!\bin{x_1\,x_2}{y_1\,x_2} = \tr_c(x_1 y_1)\,\tr_c(x_2 y_2)
+ \tr_c(x_1 y_2)\,\tr_c(x_2 y_1)\label{23}
\ee
--- the two-particle function of the free field where $\tr_c = (\free)
\,^{-1}$ is the free propagator. In accordance with the connected
structures correspondence condition (\ref{20})
$[{\cal G}^0_2]^{con}=0$ that is we have for the two-particle function
${\cal G}^0_2$:
\be
{\cal G}^0_2\!\bin{x_1\,x_2}{y_1\,y_2}=\tr^0(x_1y_1)\tr^0(x_2y_2)
+\tr^0(x_1y_2)\tr^0(x_2y_1).
\label{24}
\ee
So we have the equation for the propagator $\tr^0$ at the leading
approximation.

At the next step $(N=1)$ the system contains two equations. As it
follows from (\ref{15})
\[
(\free+\eta) * \vv{G_{pert}^1}{} + G_{pert}^1=
\left (\lambda\vv{G_{pert}^0}{}\right )\rvs
\]
so the system can be written as:
\ba
\int dx'\lambda\, (xx')\,{\cal G}^1_2\!\!\bin{x\,x'}{y\,x'}
+(\free)& & \mbox{\hspace*{-6 true mm}}\tr^1 (xy)-\delta(x-y) =
\nonumber \\
& &\mbox{\hspace{-17 true mm}} \int dx'\lambda\,
(xx')\left [g^1_2\!\!\bin{x\,x'}{y\,x'} -
g^0_2\!\!\bin{x\,x'}{y\,x'}\right ],
\label{25}
\ea
\ba
& &\mbox{\hspace*{-8 true mm}} \int dx'\lambda\,(x_1x')\,
{\cal G}^1_3\!\!\bin{x_1x_2x'}{y_1y_2x'} + (m^2-\partial\,^2_{x_1})\,
{\cal G}^1_2\!\!\bin{x_1\,x_2}{y_1\,y_2} -
\delta(x_1-y_1)\,\tr^1(x_2y_2)- \nonumber\\
& &\mbox{\hspace*{-8 true mm}} \delta (x_1 - y_2)\,\tr^1(x_2y_1)
=\int dx'\lambda\,(x_1x')
\left [g^1_3\!\!\bin{x_1\,x_2\,x'}{y_1\,y_2\,x'}
-g^0_3\!\!\bin{x_1\,x_2\,x'}{y_1\,y_2\,x'}\right].
\label{26}
\ea

At $N=1$ that is at the order $O(\lambda)$ of the perturbation theory
the connected part of the three-particle function $g^1_3$ is equal to
zero. Hence the connected structures correspondence condition at this
step reads:
\be
\left [{\cal G}^1_3\right ]^{con}=0\label{27}
\ee
and in accordance with (\ref{18}-\ref{19}) equations (\ref{25}) and
(\ref{26} ) form the system for $\tr^1$ and $\left [{\cal G}^1_2
\right]^{con}$.

To the end of this section we point out another way of the
approximation scheme construction based on reducing the order of the
Schwinger equation.

Let us recall that if any particular solution of a linear $n$-th
order differential equation is known the equation can be reduced
to that of $(n-1)$-th order. This procedure can be applied to the
Schwinger equation (\ref{10}) treated as a linear second order
differential equation with respect to the functional derivatives
\cite{3,7}.

Define a functional  $W$ in accordance with the formula
\be
W(xy|\eta)=R\,(xy|\eta)\,G[\eta]-\vv{G}{(yx)}\label{28}
\ee
where $R$ is a solution of the following equation:
\be
(\lambda\, R)\; \rvs + (\lambda R) *  R =1+(\free +\eta) * R\,.
\label{29}
\ee

The equation for $W$ follows from (\ref{10}) and (\ref{29}) and turns
out to be of the first order:
\be
(\lambda\, W)\; \rvs + (\lambda R) * W = (\free +\eta) *  W.
\label{30}
\ee

To use this method one should find at least one exact solution of
equation (\ref{29}) or equivalently the particular solution $G_{part}$
of (\ref{10}) (then $R_{part}=\vv{}{}\ln G_{part}$). Such a solution
was found in \cite{6} (see also \cite{3,7}) and is of the form:
\be
R_{part}=\int dx'dy'\lambda^{-1}(yy')\,\eta(y'x')
+m^2\int dy'\lambda^{-1}(yy').
\label{31}
\ee

On the other hand one can perform ``approximate'' lowering of the
order by using the perturbative approximant $R^N$ which is defined
as being an approximate perturbative solution of (\ref{29}) up to
$O(\lambda^N)$ accuracy.

On solving the equation for $W^N$
\be
(\lambda W^N)\; \rvs + (\lambda R^N) * W^N
=(\free +\eta) * W^N \label{32}
\ee
we obtain the sequence of approximants $\{R^N,\, W^N\}$ and from
(\ref{28}) can recover all the derivatives at $\eta=0$ of the
corresponding functional $\tilde G^N$. For the trivial solution
$W^N=0$ we get $\tilde G^N=G_{pert}^N$ --- the perturbative
approximant  (\ref{16}) and for any nontrivial $W^N\not\equiv 0$
we find some non-perturbative approximation. The problem of
supplementary boundary conditions can be solved with the help of the
connected structures correspondence condition. The final equation
for the propagator at the leading approximation coincides with
(\ref{22}).

But it should be mentioned that this way is more technically
complicated  because of the necessity to take into account nontrivial
integrability and Bose-symmetry conditions for $W$. An advantage of
this scheme is the possibility to use as a starting point the exact
solution (\ref{31}). But in so doing the question about the boundary
condition should be solved anew  because it is clear that
$R_{pert}=\lim_{N \to \infty}R^N\not=R_{part}$.
\vskip 5 true mm

\noindent
{\large \bf 2. The solution of the equation for the propagator}
\vskip 2 true mm

At the leading approximation  $(N=0)$  equation (\ref{22}) with
(\ref{23}) ¨ (\ref{24}) gives the equation for the propagator
$\tr^0$ which will be denoted as $\tr$ in this section.

Note, that at this step of the approximation we cannot use the
principle of minimal energy value of the ground state in order
to fix the physical solution because for so  doing  one should
investigate (as minimum) equations of the next step ($N=1$) in
which the connected two-particle function is contained.

Nevertheless already at the leading approximation it is possible
to solve the question of nontriviality of our scheme that is to
verify if the equation for the propagator admits solutions
different from $\tr_c$. If exist, these nontrivial solutions will
be of the nonperturbative origin by the construction.

The equation for $\tr$  can be written in the form:
\be
(\free +\Sigma) * \tr =(\free +\Sigma_c) * \tr_c\label{33}
\ee
The operators $\Sigma$ and $\Sigma_c$ are formally defined as:
\be
\Sigma (xy)\equiv \lambda (xy)\tr(xy)
+\delta(x-y)\int dx_1\lambda (xx_1)\tr (x_1x_1),\label{35}
\ee
\be
\Sigma_c (xy)\equiv \lambda (xy)\tr_c(xy)
+\delta(x-y)\int dx_1\lambda (xx_1)\tr_c (x_1x_1),\label{sc}
\ee

The structure of these quantities coincide with that of the mass
operator at the order $O(\lambda)$ of the perturbation theory. In
the local limit $\lambda (xy)\rightarrow \lambda\delta (x-y)$ they
are the formal divergent quantities and the problem arises of
equation (\ref{33}) renormalizing. Due to $\Sigma$ is expressed
in terms of the propagator to be sought  it is sufficient to define
$\Sigma_c$.  The usual way of handling such a quantity is to
introduce a regularization $\Sigma_c\rightarrow reg\Sigma_c$ and
then to define a renormalized quantity by subtractions. However
this technique is inconvenient and complicated in our case.

More simple recipe is to apply ``the renormalization without
subtraction'' method \cite{8}. This method as well as that of
``differential renormalization'' \cite{9,10} is one of the
realizations of Bogolubov's idea to  define products of
distributions without using senseless divergent quantities. The
general idea of the method is to define such a product as a
distribution from the Schwartz space being a solution of some
equation well defined in this space.

In the perturbation theory this method is equivalent to the usual
ones. The results thus obtained can differ only by finite
renormalization as they do for different subtraction procedures.

The quantity $\Sigma_c$ to be defined consist of the two terms. At
the local limit these terms are equal to each other and $\Sigma_c$
can be formally written as $2\lambda\,\tr_c(0)\delta(x-y)$. Due to
$\tr_c(x)$ has a singularity in the origin when $d\geq 2$ this
formal expression needs to be correctly defined.

Define this as a solution of an equation in the Schwartz space. The
equation can be found as follows. Let us consider the quantity:
$$
(x-y)^2\tr_c(xy).
$$
In $d\leq 4$ this is regular at $x=y$ because the singularity of the
propagator $\tr_c$ is ``killed'' by the factor $(x-y)^2$. Therefore
the expression $(x-y)^2\tr_c(xy)\lb(xy)$ is a well defined
distribution at the local limit $\lambda (xy)\rightarrow \lambda\,
\delta (x-y)$. We lay this quite a definite in sense of distributions
limit into the base of $\Sigma_c$ definition. Namely we will take
$\Sigma_c$ as a solution of the equation
$$
(x - y)^2 \Sigma_c(xy) = 2\lb \left. ((x - y)^2\tr_c(xy))
\right|_{x=y}\,\delta(x - y).
$$

i) $d=2$. In the two-dimensional space we have under $x\rightarrow y$
\be
\tr_c(xy)=-\frac{1}{4\pi}\ln
\left [\frac{(x-y)^2m^2}{4}\right ]+\psi (1)
+O\left ((x-y)^2\ln\left [(x-y)^2m^2\right ]\right ).\label{37}
\ee
If follows from (\ref{37}) that at local limit $\lambda (xy)
\rightarrow \lambda\,\delta (x-y)$ the mass operator $\Sigma_c$
satisfy the equation:
\be
(x-y)^2\Sigma_c(xy)=0.\label{38}
\ee
We will define $\Sigma_c$ as a $O(2)$-invariant solution of (\ref{38})
belonging to the Schwartz space $S'(E_2)$.
Such a solution is
\be
\Sigma_c(xy)=C\,\delta (x-y)\label{39}
\ee
where $C$ is a constant. It is a finite mass renormalization
as can be expected. From formula (\ref{37}) one can also find that
the difference of the two free propagators with masses $m_1$ and
$m_2$
\be
\tr_c(xy|m_1)-\tr_c(xy|m_2)
=\frac{1}{4\pi}\ln\frac{m^2_2}{m^2_1}
+O((x-y)^2\ln[(x-y)^2m^2])\label{40}
\ee
is non-singular at coinciding arguments. Therefore at $d=2$
\be
\Sigma_c(xy|m_1)-\Sigma_c(xy|m_2)=\frac{\lambda}{2\pi}
\,\delta(x-y)\ln\frac{m_2^2}{m^2_1}.\label{41}
\ee
That is a finite renormalization cancels in the difference of the two
mass operators. It is a reflection of the fact that the divergencies
of the corresponding integrals in the momentum space cancel.\\
\bigskip

ii) $d=3$. This case is similar to the previous one. In the three
dimensional space:
\be
\tr_c(xy)=\frac{1}{4\pi\sqrt{(x-y)^2}}
exp\{-m\sqrt{(x-y)^2}\}\label{42}
\ee
and $\Sigma_c$ is $O(3)$-invariant solution of  (\ref{38})
belonging to the Schwartz space  $S'(E_3)$ and is given by the same
formula (\ref{39}). The two mass operators difference also does not
contain the renormalization arbitrariness and reads as follows
\be
\Sigma_c(xy|m_1)-\Sigma_c(xy|m_2)
=\frac{\lambda}{2\pi}\,\delta (x-y)\,(m_2-m_1).\label{43}
\ee
Here $m\equiv\sqrt{m^2}$.
\bigskip

iii) $d=4$. In this case the theory $\phi^4$ contains quadratic
divergencies and result changes. The free propagator is more singular
at coinciding arguments
\be
\tr_c(xy)=\frac{1}{4\pi^2(x-y)^2}
+O(\ln [(x-y)^2m^2]).\label{44}
\ee
As a consequence we get an inhomogeneous equation for $\Sigma_c$
at the local limit
\be
(x-y)^2\Sigma_c(xy)=\frac{\lambda}{2\pi^2}\,\delta(x-y).\label{45}
\ee
Its $O(4)$-invariant solution from the Schwartz space $S'(E_4)$
is
\be
\Sigma_c(xy)=\frac{\lambda}{16\pi^2}\,
\partial\,^2\delta(x-y)+C\delta(x-y)\label{46}
\ee
that is $\Sigma_c$ contains the finite term of the wave function
renormalization besides the mass renormalization. This term reflects
the presence of the quadratic divergencies. In contrast to the
previous cases the difference of two $\Sigma_c$
\be
\Sigma_c(xy|m_1)-\Sigma_c(xy|m_2)=C'\delta(x-y)\label{47}
\ee
contains a finite arbitrariness. To say another, in the corresponding
momentum integrals the quadratic divergency cancels but the
logarithmic one still remains.

Having defined $\Sigma_c$ in (\ref{sc}) as described above let us
turn to the solutions of equation (\ref{33}). In what follows we will
everywhere assume $m^2>0$. When $m^2<0$ (Goldstone case) a vacuum
reconstruction is needed which leads to changing both the initial
perturbation theory and all the equations of the approximation scheme.
\bigskip

i) $d=2$. We will seek solutions for which the representation
$\Sigma(xy)=\Sigma\delta(x-y)$ is valid, $\Sigma$ being a constant.
Then equation (\ref{33}) has the following solution in the momentum
space:
\be
\tr =\frac{A}{\mu^2+p^2}+\frac{B}{m^2+p^2},\label{48}
\ee
where
\be
A=\frac{m'^2-\mu^2}{m^2-\mu^2},\qquad
B=\frac{m^2-m'^2}{m^2-\mu^2}=1-A\label{49}
\ee
and the notations are introduced:
\be
\mu^2=m^2+\Sigma,\quad m'^2=m^2+C=m^2+\Sigma_c[m].\label{50}
\ee
The operator $\Sigma$ built from $\tr$ in accordance with (\ref{sc}),
(\ref{35}) and (\ref{48}) reads
\be
\Sigma = A\Sigma_c[\mu]+B\Sigma_c[m]
= A(\Sigma_c[\mu]-\Sigma_c[m])+\Sigma_c[m].\label{51}
\ee
Taking into account formula (\ref{41}) and definition (\ref{50})
leads to an equation for $\mu^2$
\be
\frac{\lambda}{2\pi}
\frac{m'^2-\mu^2}{\mu^2-m^2}\ln\frac{m^2}{\mu^2}=m'^2-\mu^2.
\label{52}
\ee
Equation (\ref{52}) has two solutions. One of them is $\mu^2=m'^2$
which implies $A=0$, $B=1$ hence $\tr=\tr_c$. This is the trivial
solution. Another one is a solution of the equation:
\be
\frac{\lambda}{2\pi}\frac{1}{\mu^2-m^2}\ln\frac{m^2}{\mu^2}=1.
\label{53}
\ee
A solution of (\ref{53}) exists if $\lambda$ is negative and at small
$\lambda$ can be written as
\be
\mu^2\simeq m^2\exp\left(-\frac{2\pi m^2}{|\lb|}\right).\label{54}
\ee
The corresponding values of $A$ and $B$ are
\be
A\simeq \frac{m'^2}{m^2}=1+O(\lambda),\,\,
B\simeq \frac{m^2-m'^2}{m^2}=O(\lambda)\label{55}
\ee
and propagator is:
\be
\tr\simeq\frac{1}{\mu^2+p^2}+O(\lambda).\label{56}
\ee
The following circumstances connected with solution (\ref{54}
- \ref{56}) should be mentioned. First, the non-perturbative
character of the solution is obvious. Second, the negative value of
the $\lambda$ parameter does not obligatory mean that the value of
the physical coupling constant is negative. To answer the question
about the vacuum stability and the sign of physical coupling
corresponding to this solution one should at least consider the next
step of the approximation scheme when $[{\cal G}^1_2]^{con}\not=0$.
Such a situation is not unusual one in the quantum field theory
(see \cite{11}, for example).
\bigskip

ii) $d=3$. Let us try to find a solution $\Sigma(xy)=\Sigma
\delta(x-y)$ again. All formulae (\ref{48})-(\ref{51}) remain to be
valid in this case too. The only change consists in using formula
(\ref{43}) for the two mass operators difference. The equation for
$\mu^2$ has the form:
\be
m'^2-\mu^2 = - \frac{\lambda}{2\pi}\frac{m'^2-\mu^2}{m+\mu}.
\label{57}
\ee
Besides the trivial solution  $\mu^2=m'^2\,\,(\tr=\tr_c)$ the
non-perturbative one exists:
\be
\mu=-m-\frac{\lambda}{2\pi}.\label{58}
\ee
Since $\mu\equiv\sqrt{\mu^2}$ this solution exists at
$\lambda\leq-2\pi m$.
\bigskip

iii) $d=4$. Now $\Sigma_c=-\bar\lambda\, p^2+C$ where we have denoted
$$
\bar\lambda = {\lambda \over 16\pi^2}.
$$
A solution should be sought in the form $\Sigma= a p^2+b$. It still
looks as (\ref{48}) but now
\be
A=\frac{m'^2-(1-\bar\lambda)\,\mu^2}{(1+a)\,(m^2-\mu^2)},\quad
B=\frac{(1-\bar\lambda)\,m^2-m'^2}{(1+a)\,(m^2-\mu^2)},\label{59}
\ee
where
\be
\mu^2=\frac{m^2+b}{1+a},\quad m'^2=m^2+C.\label{60}
\ee
It follows from definitions (\ref{35}-\ref{sc}) and equations
(\ref{48}) and (\ref{59}) that
\ba
\Sigma&=&ap^2+b=A\Sigma_c[\mu]
+B\Sigma_c[m]=\nonumber \\
&=&  A(\Sigma_c[\mu]-\Sigma_c[m])+\frac{1-{\bar\lambda}}{1+a}
\Sigma_c[m].\label{61}
\ea
It leads to the system of equations for $a$ and $b$:
\be
\left\{\begin{array}{c}
a\,(1+a) =\bar\lambda\,(\bar\lambda -1)\\ \\
b=AC'+{\displaystyle \frac{1-\bar\lambda}{1+a}}\,C.
\end{array}\right.
\label{62}
\ee
We find the two solutions for $a$: $a_1=-\bar\lambda$ and
$a_2=\bar\lambda-1$. For the first of them $A+B=1$ and $\mu^2=m^2
+O(\lambda),\,\,a=1+O(\lambda)$. This solution is perturbatively
close to the free one:
\be
\tr=\tr_c+O(\lambda).\label{63}
\ee

The second solution gives
$$
A+B={(1-\bar\lambda) \over \mbox{\rule[3pt]{0pt}{3mm}}\bar\lambda},
\qquad
A={1 \over \mbox{\rule[3pt]{0pt}{3mm}}\bar\lambda}+O(1).
$$
This solution is  of the non-perturbative character:
\be
\tr=\frac{1}{\mbox{\rule[3pt]{0pt}{3mm}}\bar\lambda\,(\mu^2+p^2)}
+O(1).
\label{64}
\ee

Besides above mentioned solutions there can exist the ones with
a dipole term:
\be
\tr=\frac{A'}{m^2+p^2}+\frac{B'}{(m^2+p^2)^2}.\label{65}
\ee
We will not dwell upon this case which as a matter of fact is the
particular case of the previous consideration when $b\rightarrow
m^2a$. We only point out that the dipole term is always of the higher
order of the smallness in comparison with the simple pole
contribution: $B'=o_{\lb}(A')$.

To the end of this section we would like to discuss the
regularization dependence of our results. First, such a dependence
is obviously absent when $d<4$. Indeed, introducing quantities
\be
\tr' = \tr - \tr_c,\quad \Sigma' = \Sigma - \Sigma_c,\label{66}
\ee
in place of  $\tr$ and $\Sigma$, we obtain from (\ref{sc}) the
following equation:
\be
\tr' = - (m^2 - \partial\,^2+\Sigma'+\Sigma_c)^{-1} * \Sigma' * \tr_c.
\label{67}
\ee
One can see from (\ref{67}) that at $d<4$ the quantity $\tr'(x=0)$
is a finite one and therefore so is $\Sigma'$. Having solved
equation (\ref{67}) we get, of course, all the same results as we
do above, irrespective of the regularization.

In case of $d=4$ the situation is more complicated. When using the
standard regularization schemes like cutoff one we find qualitatively
the same results but the problem of cutoff removing arises. In our
approach this problem do not appear at all.
\vskip 5 true mm

\noindent
{\large \bf Conclusion}
\vskip 2 true mm

The main idea of the proposed approach is to enlarge the class of
solutions of the Schwinger equation by widening the set of admissible
boundary conditions. But the problem of boundary conditions itself
remains unsolved, of course. In essence, when constructing the
approximation scheme in section~1 we constrained ourselves to the
solutions which ``copy'' the connected structure of the perturbation
theory. But even under such hard restriction it is possible to find
the solutions of the non-perturbative character (section~2). An
appropriate choice of the solution can be made in comparing it with
the physical phenomena are to be described and on the base of general
physical principles such as the minimal energy value of the ground
state.

In this connection the investigation of the next step of the scheme
containing the two-particle amplitude (equations (\ref{25})
- (\ref{26})) is of undoubted interest. Also it would be interesting
to apply this method to other more sophisticated quantum field
models.

At last it is worth mentioning that one can construct an approximation
scheme taking as an input the loop expansion or the kinetic expansion
\cite{12} instead of the perturbation theory.

The summation of the perturbative series, independently of how far
ahead this problem could be elaborated cannot give us all the variety
of solutions of the quantum field equations. These equations contain
much more information than the sum of the perturbation theory does.
We hope that the development of approximation schemes taking into
account the singular character of the interacting fields equations
allows one to enlarge the number of the phenomena described by the
quantum field theory.
\vskip 5 true mm

\centerline{\bf Acknowledgements}
\vskip 2 true mm

We would like to thank  A.I.Alekseev, B.A.Arbuzov and V.A.Petrov for
valuable discussions and interest in this work.


\begin{thebibliography}{99}
\bibitem{1}
W. Wazow,  ``Asymptotic expansions for ordinary differential
equations'', John Wiley \& Sons publ., New York, 1965.
\bibitem{2}
A.B.Vassilieva and V.F.Butusov, ``Asymptotic expansions of
solutions of singularly perturbed equations'', Moscow,
Nauka  publ., 1973 (in Russian).
\bibitem{3}
V.E.Rochev and P.A.Saponov, ``Approximation scheme and anomalous
solutions of Schwinger equation'', talk at XI International Workshop
on High Energy Physics and Quantum Field Theory, 16--22 September
1994, Zvenigorod, Russia.
\bibitem{4}
R.J.Cant and R.J.Rivers,  J.Phys.A: Math.Gen.,{\bf 13}, 1623 (1980).
\bibitem{5}
C.M.Bender, F.Cooper, and L.M.Simmons, Phys. Rev.  D {\bf 39}, 2343
(1989).
\bibitem{6}
V.E.Rochev, J.Phys.A.: Math Gen., {\bf 26} 1235 (1993).
\bibitem{7}
V.E.Rochev, IHEP Report No. 93-107, 1993 (unpublished).
\bibitem{8}
F.A.Lunev, Phys. Rev. D {\bf 50}, 6589  (1994).
\bibitem{9}
D.Z.Freedman, K.Johnson, and J.I.Latorre, Nucl. Phys. {\bf B371} 353
(1992).
\bibitem{10}
O.I.Zavialov and V.A. Smirnov, Teor. Mat. Fiz. {\bf 96}, 288 (1993).
\bibitem{11}
P.M.Stevenson,  Phys. Rev. D {\bf 32}, 1389 (1985).
\bibitem{12}
V.E.Rochev, Nuovo Cim. {\bf 106A}, 525 (1993).
\end{thebibliography}
\end{document}